# Retrieving positions of closely packed sub-wavelength nanoparticles from their diffraction patterns

Benquan Wang[1,2], Ruyi An[3], Eng Aik Chan[1,2], Giorgio Adamo[1,2], Jin-Kyu So[1,2], Yewen Li[3],

Zexiang Shen[1,2], Bo An[3] and Nikolay I. Zheludev[1,2,4]

[1]Centre for Disruptive Photonic Technologies, TPI, Nanyang Technological University, 21 Nanyang Link, Singapore, 637371, Singapore

[2]Division of Physics and Applied Physics, School of Physical and Mathematical Sciences, Nanyang Technological University, Singapore, 637371, Singapore

[3] School of Computer Science and Engineering, Nanyang Technological University, Singapore, 639798, Singapore

[4]Centre for Photonic Metamaterials and Optoelectronics Research Centre, University of Southampton, Southampton SO17 1BJ, UK

Abstract: **Distinguishing two objects or point sources located closer than the Rayleigh distance is impossible in conventional microscopy. Understandably, the task becomes increasingly harder with a growing number of particles placed in close proximity. It has been recently demonstrated that subwavelength nanoparticles in closely packed clusters can be counted by AI-enabled analysis of the diffraction patterns of coherent light scattered by the cluster. Here we show that deep learning analysis can determine the actual position of the nanoparticle in the cluster of subwavelength particles from a sing-shot diffraction pattern even if they are separated by distances below the Rayleigh resolution limit of a conventional microscope.**

Imaging, localization, and retrieval of the number of subwavelength objects closely packed, although extremely challenging, is a problem that is very often encountered in applications such as environmental monitoring[1], semiconductor optical inspection[2], materials[3] and bio-medical analysis[4]. This problem can't be tackled by conventional microscopy, which is bound by the Abbe- Rayleigh diffraction limit to a resolution of about half the wavelength of the incident light. Improved resolution can be obtained by using optical techniques such as PALM and STED, which work with photoactivated labels [5-7] or near-field methods [8, 9], which require contact with the sample, and are therefore unsuitable in many instances because their complexity and invasiveness [10].

It was recently reported that deep learning-enabled analysis of single-shot diffraction patterns of coherent light scattered by subwavelength objects can be used to obtain unlabeled super-resolution optical metrology [11, 12] and to correctly predict the number of nano-objects in clusters of subwavelength objects [13]. Here we show that AI-empowered analysis of the optical diffraction patterns of closely packed subwavelength nanoholes, using a neural network trained on similar *a priori* known objects, allows us to retrieve their positions, even when the nanoholes are touching.

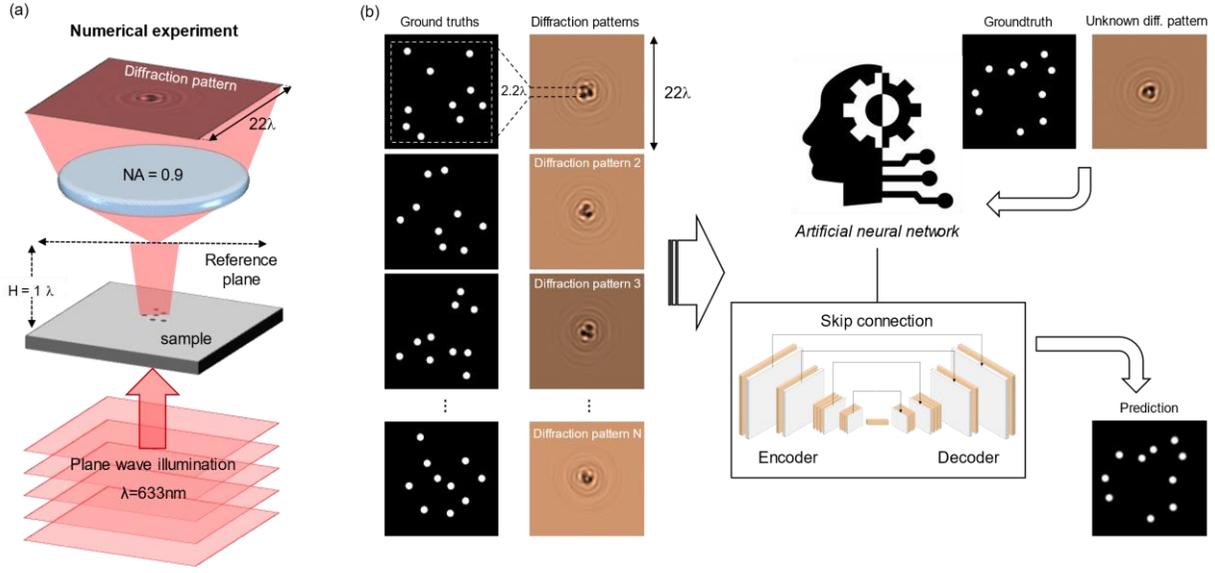

Figure 1. (a) Schematic of the numerical experiment. Clusters of subwavelength nanoholes with diameters of λ/6.33, placed within a 2.2 λ × 2.2 λ area in a 100 nm-thick chromium film, are illuminated by a plane wave with wavelength λ = 633 nm. The diffraction pattern of the scattered light intensity, at a distance, $H = 1\lambda$, is recorded with a numerical aperture, NA = 0.9. (b) Pairs of diffraction patterns and corresponding position maps of the nanoholes (ground truths) are used to train a modified U-Net, encoder-decoder convolutional neural network, which will then be able to retrieve the positions of nanoholes from single-shot unknown diffraction patterns.

We conducted numerical experiments using a coherent plane-wave illumination (λ = 633 nm) of clusters of subwavelength nanoholes with a diameter of λ/6.33 perforated in an opaque film, randomly placed within a 2.2 λ × 2.2 λ area. We image the diffraction patterns created by the nanoholes clusters at a distance, H = 1 λ away from the sample, over a 22 λ × 22 λ field of view, accounting for the numerical aperture, NA = 0.9 of a real imaging system (Fig. 1a). Pairs of the far-field diffraction pattern and the corresponding position map of nanoholes in the cluster were used to train a modified U-Net encoder-decoder convolutional neural network. U-Net is a convolutional neural network (CNN) architecture specifically designed for semantic image segmentation[14], i.e., the categorization of each pixel in an image into a class or object. The network has a U-shaped architecture, which consists of an encoder, or dimension-reducing path, followed by a decoder or dimension-increasing path, with a symmetric design that reduces the risk of information loss during the encoding and decoding processes [15]. The encoder path captures and condenses information of inputs at multiple levels of abstraction through convolutional and pooling layers, as in traditional CNNs [16]. The decoder path, instead, uses transposed convolutions for recovering the dimension, conditioned by skip connection from the correspondingly encoded information at the same level. This approach enables the network to produce precise segmentation masks and effectively addresses the vanishing gradient problem[17]. This phenomenon, where the network loses its capacity to capture long-term dependencies, is mitigated through this design. To promote efficient information propagation and resolve the category imbalance challenge (i.e., the small fraction of nanoholes with respect to the background), we modify the U-Net by introducing a residual architecture[18] and novel hybrid loss function that combines Binary Cross-Entropy loss and a Class-Balanced Loss. The trained network is then able to retrieve the positions of the particle in the cluster from previously unseen diffraction patterns (Fig.1b).

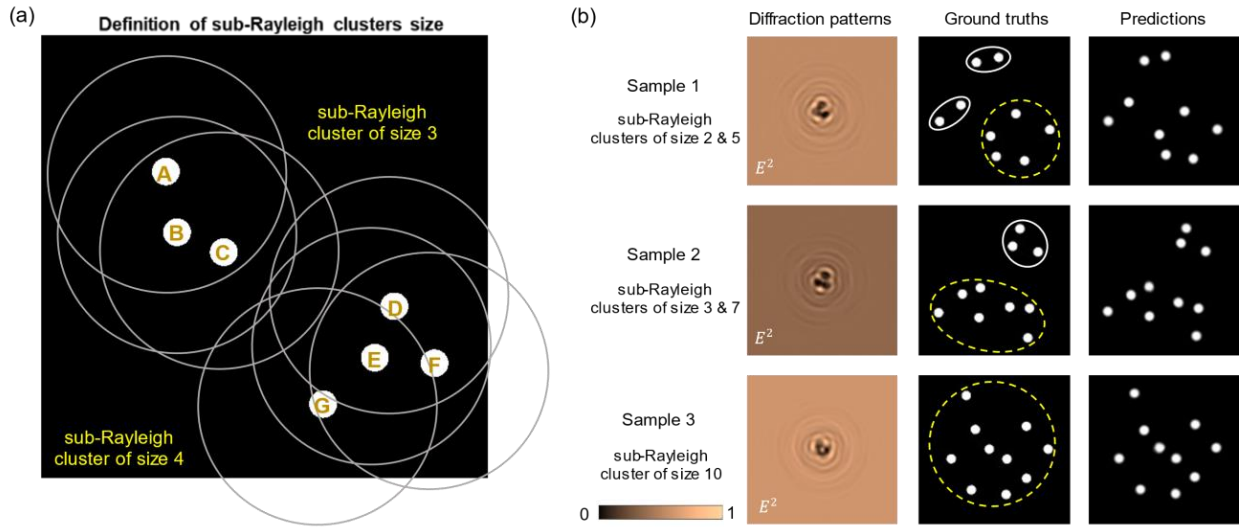

Figure 2. (a) A group of nanoholes is identified as a sub-Rayleigh cluster if each nanohole in the cluster has at least one neighboring nanohole within Rayleigh distance, 0.61 λ/NA. The binary map shown here contains a sub-Rayleigh cluster of 3 nanoholes (A, B, C) and a cluster of 4 nanoholes (D, E, F, G). The circles represent the Rayleigh region of each nanohole. Nanoholes (A, B, C) and nanoholes (D, E, F) all fall within the Rayleigh distance. (b) Diffraction patterns (first column), groundtruth (second column) and prediction (third column) images of three samples where the size of the largest Rayleigh cluster (yellow dashed circle) increases from 5 (first row), to 7 (second row) and 10 (third row).

Each sample contains up to 10 nanoholes that may form clusters with an inter-particle distance smaller than the Rayleigh limit of resolution of a conventional microscope, 0.61 λ/NA. We define the sizes of the sub-Rayleigh clusters by counting the number of nanoholes whose inter-particle distance is smaller than the Rayleigh limit (Fig. 2a) and characterize each sample by the largest sub-Rayleigh cluster size within the 2.2λ x 2.2λ area. It shall be noted how, often, not only pairs, but all particles fall within the Rayleigh distance (e.g., nanoholes A, B, C in Fig 2a). The groundtruth maps used to supervise the network are binary images of 512×512 pixels size (corresponding to an area of 2.5 λ × 2.5 λ) where white pixels (value = 1) represent the nanoholes and black pixels (value = 0) represent the Cr film (second column in Fig. 2b). The corresponding diffraction patterns were generated plotting the total electric field intensity profiles calculated by finite-difference-time-domain (FDTD) full Maxwell solver, Lumerical, at a distance of 1λ from the sample surface, over a field of view of 22λ x 22λ (first column in Fig. 2b). Fig. 2b shows that the light propagating through the nanoholes, generated very reach interference patterns in the diffraction maps, thus making it very difficult to correlate to a specific number and distribution of nanoholes on the sample. Nonetheless, the trained network can not only retrieve the number and positions of the nanoholes in the clusters but also return 512×512 pixels images (third column in Fig. 2b) where the sizes of the nanoholes match well those of the groundtruth and therefore can be regarded as a form of super resolution imaging.

A total of 11,700 samples and corresponding diffraction patterns were generated for the numerical experiment, of which 7,200 were used for training, 1800 for validation and 2,700 for test of the neural network. We use the Pearson correlation coefficient[19] between the predicted and ground truth images to evaluate the accuracy of image reconstruction of our technique.

$$r_{xy} = \frac{\sum_{i=1}^{n}(x_i - \bar{x})(y_i - \bar{y})}{\sqrt{\sum_{i=1}^{n}(x_i - \bar{x})^2}\sqrt{\sum_{i=1}^{n}(y_i - \bar{y})^2}} \quad (1)$$

where $n$ is the sample size, $x_i$, $y_i$ are the individual sample points in our reconstructed image and ground truth image, respectively, indexed with $i$, $\bar{x} = \frac{1}{n}\sum_{i=1}^{n} x_i$ (the sample mean), and analogously for $\bar{y}$.

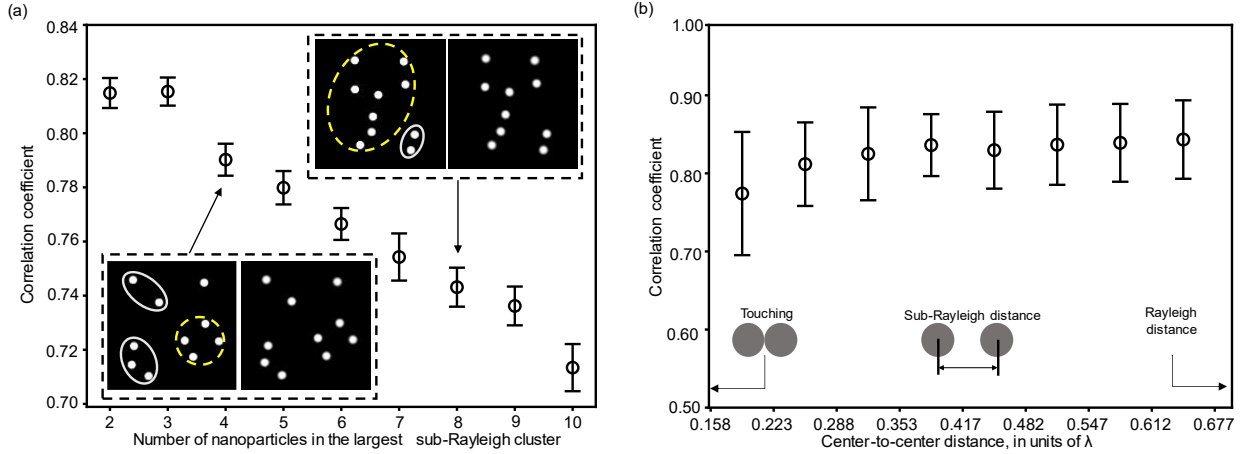

Figure 3. (a) Pearson correlation coefficient between the predicted and ground truth images as a function of the sub-Rayleigh cluster size. Insets: examples of groundtruth (left) and prediction (right) images for samples with the largest sub-Rayleigh cluster size of 4 (lower left) and 8 (upper right) nanoholes. (b) Pearson correlation coefficient between the predicted and ground truth images for the typical Rayleigh diffraction case, two closely spaced nanoholes of decreasing center-to-center distance.

Fig. 3a shows that the accuracy exceeds 0.81 for samples with the sub-Rayleigh cluster index of 3 and remains as high as 0.71 for samples containing 10 nanoholes within the same sub-Rayleigh cluster, which is an indication of the trained network's robustness against the size of sub-Rayleigh clusters. The decrement of the image reconstruction accuracy with the increasing size of sub-Rayleigh clusters can be justified by the increase in the complexity of the interference patterns.

To explore further our technique and test its resilience against the problem of closely paced particles, we tested its performance with an example of two closely spaced nanoholes. In this case, we use 800 diffraction patterns of two λ/6.33 nanoholes with center-to-center separation decreasing from 0.677λ (Rayleigh distance) to 0.158λ (touching) as a test. The Pearson correlation coefficient calculated in Fig. 3b shows that the two nanoholes could be resolved with an accuracy higher than 0.8 across the almost entire range of sub-Rayleigh distances, with a slight drop to 0.75 in the only case of touching nanoholes.

In conclusion, we report on a far-field, single-shot super-resolution optical technique based on the deep learning of the light diffracted on the clusters of subwavelength particles. It allows retrieving maps showing the number, positions, and sizes of the nanoparticles in the cluster and therefore constitutes a form of imaging. The image reconstruction accuracy measured as the correlation coefficient between the ground truth and reconstructed maps of the nanoparticles depends on the number of nanoparticles in the largest cluster of sub-Rayleigh spaced particles and varies from

0.82 to 0.71 when the size of the cluster increases from 2 to 10. In addition, we showed that the technique resolves nanoholes separated significantly smaller than the Rayleigh distance.

**References**

1. Dincer, C., et al., *Disposable sensors in diagnostics, food, and environmental monitoring.* Advanced Materials, 2019. **31**(30): p. 1806739.
2. van der Walle, P., et al. *Deep sub-wavelength metrology for advanced defect classification.* SPIE.
3. Pujals, S., et al., *Super-resolution microscopy as a powerful tool to study complex synthetic materials.* Nature Reviews Chemistry, 2019. **3**(2): p. 68-84.
4. Schermelleh, L., et al., *Super-resolution microscopy demystified.* Nature cell biology, 2019. **21**(1): p. 72-84.
5. Hell, S.W. and J. Wichmann, *Breaking the diffraction resolution limit by stimulated emission: stimulated-emission-depletion fluorescence microscopy.* Optics letters, 1994. **19**(11): p. 780-782.
6. Rust, M.J., M. Bates, and X. Zhuang, *Sub-diffraction-limit imaging by stochastic optical reconstruction microscopy (STORM).* Nature methods, 2006. **3**(10): p. 793-796.
7. Betzig, E., et al., *Imaging intracellular fluorescent proteins at nanometer resolution.* science, 2006. **313**(5793): p. 1642-1645.
8. Jiang, R.-H., et al., *Near-field plasmonic probe with super resolution and high throughput and signal-to-noise ratio.* Nano Letters, 2018. **18**(2): p. 881-885.
9. Tsai, D.P. and W.C. Lin, *Probing the near fields of the super-resolution near-field optical structure.* Applied Physics Letters, 2000. **77**(10): p. 1413-1415.
10. Astratov, V.N., et al., *Roadmap on Label-Free Super-Resolution Imaging.* Laser & Photonics Reviews, 2023.
11. Rendón-Barraza, C., et al., *Deeply sub-wavelength non-contact optical metrology of sub-wavelength objects.* APL Photonics, 2021. **6**(6).
12. Pu, T., et al., *Unlabeled Far-Field Deeply Subwavelength Topological Microscopy (DSTM).* Adv Sci (Weinh), 2020. **8**(1): p. 2002886.
13. Chan, E.A., et al., *Counting and mapping of subwavelength nanoparticles from a single shot scattering pattern.* Nanophotonics, 2023.
14. Ronneberger, O., P. Fischer, and T. Brox. *U-net: Convolutional networks for biomedical image segmentation.* in *Medical Image Computing and Computer-Assisted Intervention–MICCAI 2015: 18th International Conference, Munich, Germany, October 5-9, 2015, Proceedings, Part III 18.* 2015. Springer.
15. Zhang, W., et al., *High-axial-resolution single-molecule localization under dense excitation with a multi-channel deep U-Net.* Optics Letters, 2021. **46**(21): p. 5477-5480.
16. Gu, J., et al., *Recent advances in convolutional neural networks.* Pattern recognition, 2018. **77**: p. 354-377.
17. Hochreiter, S., *The vanishing gradient problem during learning recurrent neural nets and problem solutions.* International Journal of Uncertainty, Fuzziness and Knowledge-Based Systems, 1998. **6**(02): p. 107-116.
18. He, K., et al. *Deep residual learning for image recognition.* in *Proceedings of the IEEE conference on computer vision and pattern recognition.* 2016.

19. Rice, J.A., *Mathematical statistics and data analysis*. 3rd ed. Duxbury advanced series. 2007, Belmont, CA: Thomson/Brooks/Cole.

**Acknowledgments:** This work was supported by the Singapore National Research Foundation (Grant No. NRF-CRP23-2019-0006), the Singapore Ministry of Education (Grant No. MOE2016-T3-1-006), and the Engineering and Physical Sciences Research Council UK (Grants No. EP/T02643X/1).